\documentclass{sig-alternate-05-2015}
\usepackage{colortbl}
\usepackage{color}
\usepackage{url}
\usepackage{multirow}
\usepackage{arydshln}
\usepackage{pgf}
\usepackage{tikz}
\usepackage{pgfplots}
\usepackage{subfigure}
\definecolor{light-gray}{gray}{0.80}

\begin{document}

\CopyrightYear{2016} 
\setcopyright{acmcopyright}
\conferenceinfo{BDCAT'16,}{December 06-09, 2016, Shanghai, China}
\isbn{978-1-4503-4617-7/16/12}
\acmPrice{\$15.00}
\doi{http://dx.doi.org/10.1145/3006299.3006315}

%

\title{A Study of Factuality, Objectivity and Relevance: \\Three Desiderata in Large-Scale Information Retrieval?}
%
%
%
%
%

\numberofauthors{2} 
%
\author{
%
%
\alignauthor
Christina Lioma\\
       \affaddr{Department of Computer Science}\\
       \affaddr{University of Copenhagen}\\
       \affaddr{Universitetsparken 5, 2100 Copenhagen, Denmark}\\
       \email{c.lioma@di.ku.dk}
\alignauthor
Birger Larsen\\
       \affaddr{Department of Communication}\\
       \affaddr{Aalborg University in Copenhagen}\\
       \affaddr{A.C. Meyers Vaenge 15, 2450 Copenhagen, Denmark}\\
       \email{birger@hum.aau.dk}
\and
\alignauthor Wei Lu\\
       \affaddr{School of Information Management}\\
       \affaddr{Wuhan University}\\
       \affaddr{No. 299 Bayi Road, Wuhan, China}\\
       \email{weilu@whu.edu.cn}
\alignauthor Yong Huang\\
       \affaddr{School of Information Management}\\
       \affaddr{Wuhan University}\\
       \affaddr{No. 299 Bayi Road, Wuhan, China}\\
       \email{huangyng@gmail.com}
}

\maketitle
\begin{abstract}
Much of the information processed by Information Retrieval (IR) systems is unreliable, biased, and generally untrustworthy \cite{Fallows05,Metzger07,MorrisCRHS12}. Yet, factuality \& objectivity detection is not a standard component of IR systems, even though it has been possible in Natural Language Processing (NLP) in the last decade. Motivated by this, we ask if and how factuality \& objectivity detection may benefit IR. 
We answer this in two parts. First, we use state-of-the-art NLP to compute the probability of document factuality \& objectivity in two TREC collections, and analyse its relation to document relevance. We find that factuality is strongly and positively correlated to document relevance, but objectivity is not. Second, we study the impact of factuality \& objectivity to retrieval effectiveness by treating them as query independent features that we combine with a competitive language modelling baseline. Experiments with 450 TREC queries show that factuality improves precision by more than 10\% over strong baselines, especially for the type of uncurated data typically used in web search; objectivity gives mixed results. An overall clear trend is that document factuality \& objectivity is much more beneficial to IR when searching uncurated (e.g. web) documents vs. curated (e.g. state documentation and newswire articles). 

To our knowledge, this is the first study of factuality \& objectivity for back-end IR, contributing novel findings about the relation between relevance and factuality/objectivity, and statistically significant gains to retrieval effectiveness in the competitive web search task.
\end{abstract}

%
%


%
%

%
%



\section{Introduction}
\label{s:intro}

Information \textit{factuality} is typically understood as the extent to which accounts, opinions, evaluations or emotions that vary across multiple reporters are factual \cite{Froehlich02}. Information is factual, hence reliable, if one can accept it as the truth without needing to check its validity elsewhere \cite{SchwarzM11}.
Information \textit{objectivity} on the other hand, is almost always defined as the opposite of subjectivity, which is generally understood  as the semantic orientation or valence of text \cite{Hatz97,OsgoodST57}, or more simply the extent to which text meaning depends on the author's perspective \cite{Hill14}. Objectivity is the degree to which language is interpretable independently of the speaker's perspective \cite{Lang02}. Factuality and objectivity are well established cognitive and linguistic notions \cite{Hill14}, both often manifested when opinions or evaluations are conveyed.

Intuitively, retrieving unreliable and subjective information seems undesirable, as it can impact negatively decision making \cite{Froehlich02} and may require time-consuming fact cross-validation from other sources \cite{SchwarzM11}. However, whereas assessing factuality \& objectivity seems a standard cognitive task for humans \cite{Can13}, search engine users often mistake the output of web search engines as a confirmation of factuality or objectivity. For instance, many interpret the ranking of search results as a key indicator of credibility \cite{Hargittai10}, whereas social network users often determine the truthfulness of content based on misleading heuristics, such as if an item is re-tweeted \cite{MorrisCRHS12}. In fact, in 2005, a majority of U.S. web users considered search engines fair and unbiased as an information source \cite{Fallows05}. This is not always the case; much of the information on the web is actually unreliable, biased, and generally untrustworthy \cite{Fallows05,Metzger07,MorrisCRHS12}.

As a result of this, IR research has started addressing aspects of factuality \& objectivity (reviewed in Section \ref{s:rw}). 
NLP research has been addressing the detection and measurement of factuality \& objectivity for a longer time, with significant recent advances and resulting gains for tasks like question answering  \cite{Oh13} or information extraction \cite{WiebeR11}. 

Motivated by the need to distinguish between retrieving credible over non-credible information, and by the NLP advances that allow its detection, we ask if and how document factuality \& objectivity can benefit IR. We choose factuality \& objectivity because they are two core dimensions of credibility \cite{Froehlich02,OsgoodST57}. We use two state-of-the-art, scalable NLP methods to estimate document factuality \& objectivity in two TREC collections (Section \ref{ss:relsub}). Statistical analysis of the collections (Section \ref{s:relobj}) and subsequent experimental evaluation in a reranking scenario (Section \ref{s:eval}) show that (i) considering the factuality of the retrieved documents improves retrieval effectiveness significantly for uncurated (e.g. web) collections but not for curated collections (e.g. older and `cleaner' TREC disks); (ii) document objectivity does not seem to improve retrieval effectiveness. 
This work contributes novel findings about the relation between relevance and factuality/objectivity, and statistically significant gains to retrieval effectiveness in the competitive web search task.

\section{Related Work}
\label{s:rw}

Previous research has investigated aspects of information factuality \& objectivity in search scenarios. 
One example is Dispute Finder \cite{EnnalsBAR10}, a system that identifies contentious topics by looking for disputed text: a factual claim is considered disputed if a webpage suggests both that the claim is false and also that other people say it is true. Dispute Finder extracts disputed claims by searching the web for patterns such as \texttt{falsely claimed that X} and using them to train a classifier to select text that makes a disputed claim. Similar to this is the \textit{factual density} measure \cite{Lex12}, which estimates factuality as the ratio of facts in a document over document length. \textit{Factual density} has been used to identify highly factual (hence reliable) articles in Wikipedia \cite{Lex12} and other web sources \cite{HornZGKL13}. Visualisations on the user interface have also been shown to help users assess information as factual, biased or opinionated \cite{GamonBBFHK08,MorrisCRHS12,Park09}. E.g., visualising the Wikipedia edit history can change users' perceptions of trustworthiness \cite{KitturSC08}; similarly, visualizing popularity among experts next to the search results can help users make more accurate credibility ratings \cite{SchwarzM11}.

Several text quality metrics exist for capturing aspects of factuality \& objectivity, typically using as features stylometric indicators of text quality \cite{LexJG10}, for instance character trigram distributions \cite{LipkaS10}, word counts \cite{Blumenstock08}, or text structure (for blogs) \cite{JuffingerGL09}. Some approaches are so far applied solely to curated data, like Wikipedia \cite{LexJG10,LipkaS10}, which is relatively homogeneous (format- and structure-wise). Processing uncurated web data requires more complex features, such as semantic relations \cite{Etzioni08}, 
extracted or inferred about entities named in text \cite{Lex12}, using as clues individual words and/or lexico-syntactic patterns \cite{WiebeR11}. The above approaches are developed stand-alone, not as an integral part of IR systems. However, methods for potentially integrating such text quality metrics into IR systems abound, for instance when text quality is interpreted as ratios of (combinations of) stopwords over content words per document \cite{BenderskyCD11, ZhouC05,ZhuG00}; term length \cite{KanungoO09} or part-of-speech (for ranking \cite{LiomaO07, LiomaK:2008}, but also for index pruning \cite{LiomaO07}); technical \cite{LarsenL12} or ambiguous scientific terminology \cite{LiomaKS11}; non-compositional multiword expressions \cite{LiomaSLH15,MichelbacherKFLS11}; document readability \cite{ColemanL75, Gunning52, KincaidFRC75, McClure87, McLaughlin69}; document discourse \cite{Lioma12} or coherence \cite{petersen, LiomaTSPL16}.

Factuality and objectivity analysis has been studied in NLP for over a decade, producing annotated corpora \cite{wiebe05}, schemes for manual annotation \cite{Herzig11}, tools for automatic analysis \cite{WilsonHSKWCCRP05}, and generally a good understanding of their computational treatment \cite{ShanahanQW06}. This has led to successful applications of factuality \& objectivity analysis to sentiment analysis \cite{Yes10}, information extraction \cite{WiebeR11}, and question answering \cite{Oh13}, 
in English but also other languages \cite{BaneaMW14}. In these tasks, the analysis is typically done at a phrase, sentence, or document level; however, more recent analysis is done at a lexical level, e.g. to disambiguate words according to their usage or sense \cite{BaneaMW14}, or for lexical semantics research \textit{per se}, e.g. showing that objective adjectives are most likely to modify concrete nouns \cite{Hill14}.

In this work, we measure factuality \& objectivity on a document level, so that we can study its relation and potential usefulness to retrieval. None of the above methods have been applied to back-end retrieval to our knowledge.

\section{Factuality and Objectivity Estimation}
\label{ss:relsub}

The first step in studying factuality \& objectivity for IR is detecting them in documents. We do this using two state-of-the-art, scalable approaches. We present these next and their application to IR documents in Section \ref{s:relobj}. 

\subsection{Detecting Factuality}
\label{sss:rel}
We estimate document factuality using \textit{factual density} \cite{Lex12}. This consists of (i) identifying factual and non-factual documents in some small corpus, (ii) using features of these documents to train a classifier, and (iii) applying the classifier to our data to estimate the factuality of each document. \textit{Factual density} uses open information extraction (Open IE) methods to extract relational tuples of facts from text. 
A relational tuple is typically a triplet of two noun phrase arguments and a relational verb phrase, e.g. $f$ \texttt{= (Poe, was born in, Boston)}, where $f$ denotes a fact. Even though Open IE is scalable to large sets of relations and corpora \cite{Etzioni08}, it is prone to extracting noise \cite{Fader11}, risking to overestimate \textit{factual density}. To avoid this, we extract facts using ReVerb, a competitive IE system that uses part-of-speech (POS) and lexical constraints on verb relations, with a lower noise extraction rate in web documents \cite{Fader11}. For implementation details see \cite{Fader11}. 

Following \cite{Lex12}, we collect randomly 2000 \textit{featured}\footnote{\url{http://en.wikipedia.org/wiki/Wikipedia:Featured_articles}} or \textit{good}\footnote{\url{http://en.wikipedia.org/wiki/Wikipedia:Good_articles}} Wikipedia articles (as positive examples of factuality), and 2000 articles from the remaining Wikipedia (as negative examples of factuality). 
\textit{Featured} and \textit{good} articles are the best articles in Wikipedia, as decided by editorial review according to accuracy, neutrality, completeness, and style, so they are suitable examples of factuality. 
Even though there is no guarantee that there will not be factual documents among our 2000 negative examples, it is likely that negative examples will be \textit{less} factual than \textit{featured/good} articles. This premise worked well for \cite{Lex12} and we therefore also adopt it. We extract facts from the 4000 positive and negative examples using ReVerb. In 60 of the negative examples, we find no facts; this supports the assumption that \textit{non-featured/non-good} articles are likely to be less factual. We use the word count, fact count, fact density (ratio of fact count over word count), and top 10\% information gain relational tuples (600 relational tuples) to train a classifier (SVM, RBF kernel) using LIBSVM and 5-fold cross validation. 
This yields 90.3\% accuracy. Using this trained model on our data outputs a probability of factuality for each document. We use this probability as is, without calibrating it. For potentially more refined results, we could have used any among several probability calibration methods specifically designed for SVM classifiers, based e.g. on sigmoid \cite{platt99}, or Bayesian transformations \cite{Kwok99}. We do not do so, given the high classification accuracy obtained, and also to have more control over the effect of factuality on retrieval; extra smoothing might fetch even better results, but with the risk of washing away other effects that might make our interpretation of the impact of factuality on relevance less transparent. 

\begin{table}
\centering
\caption{\label{tab:extraction_pattern_templates}Syntactic templates for extraction patterns used in objectivity detection (from \cite{WiebeR11}): $subj$ (subject), $dobj$ (direct object), $np$ (noun phrase), $aux$ (auxiliary verb), $prep$ (preposition).}
\scalebox{0.85}{
\begin{tabular}{l|l|l} 
$[$subj$]$ passive-verb		&active-verb $[$dobj$]$		&noun prep $[$np$]$\\
$[$subj$]$ active-verb		&infinitive $[$dobj$]$			&active-verb prep $[$np$]$\\
$[$subj$]$ active-verb dobj	&verb infitive $[$dobj$]$		&passive-verb prep $[$np$]$\\
$[$subj$]$ verb infinitive		&noun aux $[$dobj$]$		&infitive prep $[$np$]$\\
$[$subj$]$ aux noun			&						&\\
\end{tabular}
}
\end{table}

\subsection{Detecting Objectivity}
\label{sss:sub}
We estimate document objectivity with the subjectivity detection approach of Wiebe \& Riloff \cite{WiebeR11}. 
The required inputs are a subjectivity lexicon, a small set of seed nouns, and some human review, which we explain below. For implementation details see \cite{WiebeR11}.

We split our data into sentences with openNLP, and parse sentences to get typed dependencies (e.g. what is the subject or object of the sentence) with the Stanford parser\footnote{\url{http://nlp.stanford.edu/software/lex-parser.shtml}}. We filter the sentences using the extraction pattern templates of \cite{WiebeR11} shown in Table \ref{tab:extraction_pattern_templates}, and keep only sentences containing these patterns. We use the top 20 most subjective nouns in our data (manually compiled by examining the words with the highest collection term frequency that are nouns and choosing the 20 most subjective) as seeds and run 400 iterations of Meta-Bootstrapping \cite{RiloffJ99} and Basilisk \cite{Thelen:2002} on the extracted sentences. We select five new subjective nouns after each iteration and manually annotate the resulting 4000 nouns as strongly/weakly subjective or objective. 
We add these annotated nouns to an existing subjectivity lexicon of $>$8000 entries made available in \cite{WiebeR11}, and use them as features to classify sentences as subjective or objective in our dataset. A sentence is subjective if it contains $\ge2$ strongly subjective terms, and objective if it contains no strongly subjective terms and maximum 2 weakly subjective terms (as per \cite{WiebeR11}). We thus build 10000 training examples (5000 subjective and 5000 objective), which we feed into an extraction pattern learning algorithm that extracts subjective and objective patterns, such as \texttt{$[subj]$ complained}, or  \texttt{to condemn $[dobj]$} (see Table \ref{tab:extraction_pattern_templates} for abbreviations). We rank the extracted patterns according to the probability that a sentence is subjective or objective given that a specific pattern appears in it, as described in \cite{WiebeR11}, using the exact same thresholds.  

We use the above subjective/objective nouns, lexicon, patterns, plus POS features, to train an objectivity classifier (SVM, RBF kernel) using LIBSVM and 5-fold cross validation, on our 10000 training sentences. 
This yields 87.83\% accuracy. We use this trained model to estimate the probability of objectivity of each document in our data as a fraction of the number of objective sentences in it over its total number of sentences.

\section{Statistical Analysis of Factuality \& Objectivity vs. Relevance}
\label{s:relobj}
To find out if factuality \& objectivity are useful to IR we study their relation to relevance. We start with a small-scale analysis of 227,307 documents, for which we have scores of relevance, factuality \& objectivity (explained below). 

We look at the association between relevance and factuality/objectivity on two datasets of different domains: (a) a curated, `clean' newswire/official documentation corpus (TREC Disks4-5), and (b) a bigger, uncurated, relatively noisy web crawl, TREC ClueWeb09B (see Table \ref{tab:datasets}). Disks4-5 contain documents from the 103$^{rd}$ Congressional Record, the 1994 Federal Register, the 1992-1994 Financial Times, the 1996 Foreign Broadcast Information Service, and the 1989-1990 Los Angeles Times. We consider these curated, because they have been very likely professionally edited. ClueWeb09B contains webpages crawled in 2009 from heterogeneous web sources that are not necessarily professionally edited and include spam; hence, we consider this data uncurated. We select these datasets intentionally to study the effect of curated vs. uncurated data on factuality \& objectivity detection for IR. Both datasets come with TREC queries and binary relevance assessments. We study the relation between relevance and factuality/objectivity on a subset of these datasets which includes \textit{all and only} those documents for which we have TREC relevance assessments: these are $\sim$144K documents or 25.9\% of Disks4-5, and $\sim$83K documents or 0.17\% of ClueWeb09B (see Table \ref{tab:datasets}). We estimate the factuality \& objectivity of each document in these subsets as described in Section \ref{ss:relsub}, and analyse their distribution (Sec \ref{ss:relobjdist}) and their covariation with document relevance (Section \ref{ss:relobjrlv}). 

\begin{table}
\centering
\caption{\label{tab:datasets}Datasets. \# means `number of'.}
\scalebox{0.85}{
\begin{tabular}{l|c|c} 
\rowcolor{light-gray}
						&CURATED		&UNCURATED\\
						&Disks4-5			&ClueWeb09B\\
\hline
\# Documents				&556,077			&50,220,423\\
\hline
\# Queries					&301-450			&1-200\\
\hline
\multirow{2}{*}{\# TREC-assessed Documents}
						&144,144			&83,163\\
						&(25.9\%)			&(0.17\%)\\
\hline
TREC track				&AdHoc			&Web AdHoc\\
\end{tabular}
}
\end{table}

\begin{figure*}
\centering
\pgfplotsset{every-axis label/.append style={font=\LARGE}}
\tikzset{every mark/.append style={font=\HUGE}}
\scalebox{0.8}{
\begin{tabular}{cccc}
\begin{tikzpicture}[baseline,scale=0.5]
			\begin{axis}[
			align =center, 
			font=\LARGE,			
scale only axis,
scaled y ticks = false,
title={Disks4-5\\(144,144 documents)},
xlabel=p(factuality)
]
\pgfplotstableread{d45.RLB.plot}\table 
\addplot+[only marks, mark=+] table[y index=0,x index=1] from \table;
\end{axis}
\end{tikzpicture}&
%
\begin{tikzpicture}[baseline,scale=0.5]
			\begin{axis}[
			font=\LARGE,	
			align =center, 		
			scale only axis,
scaled y ticks = false,
y tick label style={/pgf/number format/fixed},
title={ClueWeb09B\\(83,163 documents)},
xlabel=p(factuality)
]
\pgfplotstableread{cweb.RLB.plot}\table 
\addplot+[only marks, mark=+] table[y index=0,x index=1] from \table;
\end{axis}
\end{tikzpicture}&
%
\begin{tikzpicture}[baseline,scale=0.5]
			\begin{axis}[
			font=\LARGE,	
			align =center, 		
			scale only axis,
scaled y ticks = false,
y tick label style={/pgf/number format/fixed},
xlabel=p(objectivity),
title={Disks4-5\\(144,144 documents)}
]
\pgfplotstableread{d45.OBJ.plot}\table 
\addplot+[only marks, mark=+] table[y index=0,x index=1] from \table;
\end{axis}
\end{tikzpicture}&
%
\begin{tikzpicture}[baseline,scale=0.5]
			\begin{axis}[
			font=\LARGE,	
			align =center, 		
			scale only axis,
			scaled y ticks = false,
y tick label style={/pgf/number format/fixed},
xlabel=p(objectivity),
title={ClueWeb09B\\(83,163 documents)}
]
\pgfplotstableread{cweb.OBJ.plot}\table 
\addplot+[only marks, mark=+] table[y index=0,x index=1] from \table;
\end{axis}
\end{tikzpicture}
\end{tabular}
}
\caption{\label{fig:dist} Distribution of probability of factuality/objectivity (binned). The y-axis shows the number of documents.}
\end{figure*}
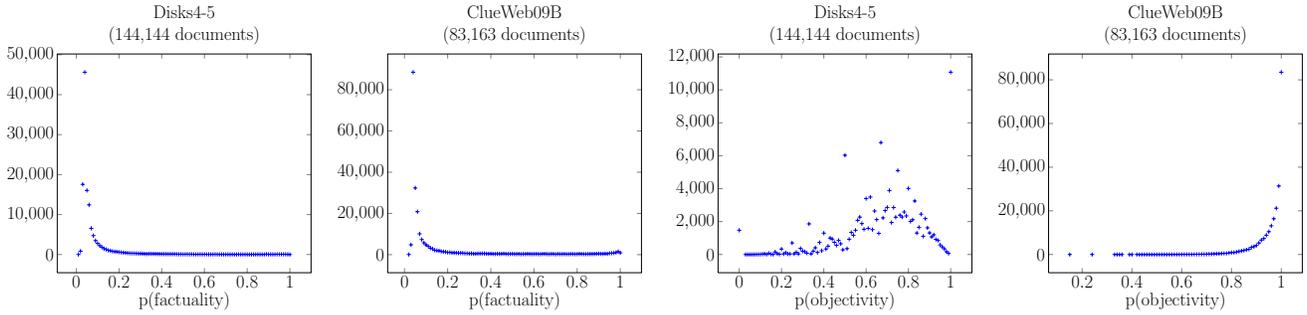

\begin{figure*}
\centering
\pgfplotsset{every-axis label/.append style={font=\LARGE}}
\tikzset{every mark/.append style={font=\huge}}
\scalebox{0.9}{
\begin{tabular}{cccc}
\begin{tikzpicture}[baseline,scale=0.5]
			\begin{axis}[
			align =center, 
			font=\LARGE,			
			title= {Disks4-5\\Spearman's $\rho$: 0.83 ($P$-value: 0)},
			ylabel=p(factuality),
			xlabel=p(relevance)
			]
			\pgfplotstableread{ProbRelev25BinReliab.plot}\table 
			\addplot[mark=+] table[x index=0,y index=1] from \table;
\end{axis}
\end{tikzpicture}&
%
\begin{tikzpicture}[baseline,scale=0.5]
			\begin{axis}[
			font=\LARGE,
			align =center, 			
			title= {ClueWeb09B\\ Spearman's $\rho$: 0.92 ($P$-value: 0)},
			ylabel=p(factuality),
			xlabel=p(relevance)
			]
			\pgfplotstableread{ProbRelev48BinReliab.plot}\table 
			\addplot[mark=+] table[x index=0,y index=1] from \table;
\end{axis}
\end{tikzpicture}&
\begin{tikzpicture}[baseline,scale=0.5]
			\begin{axis}[
			font=\LARGE,	
			align =center, 		
			title= {Disks4-5\\ Spearman's $\rho$: 0.14 ($P$-value: 0.54)},
			ylabel=p(objectivity),
			xlabel=p(relevance)
			]
			\pgfplotstableread{ProbRelev25BinObj.plot}\table 
			\addplot[mark=+] table[x index=0,y index=1] from \table;
\end{axis}
\end{tikzpicture}&

\begin{tikzpicture}[baseline,scale=0.5]
			\begin{axis}[
			font=\LARGE,
			align =center, 			
			title= {ClueWeb09B\\ Spearman's $\rho$: -0.68 ($P$-value: 0)},
			ylabel=p(objectivity),
			xlabel=p(relevance)
			]
			\pgfplotstableread{ProbRelev48BinObj.plot}\table 
			\addplot[mark=+] table[x index=0,y index=1] from \table;
\end{axis}
\end{tikzpicture}
\end{tabular}%
}
\caption{Probability of factuality \& objectivity (y-axis) vs. probability of relevance (x-axis) (binned).}
\label{fig:prelAnal}
\end{figure*}
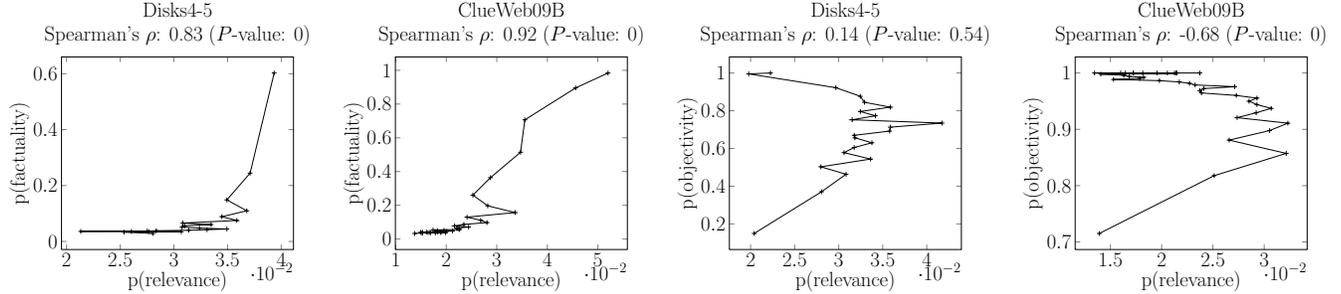

\subsection{Factuality \& Objectivity Distribution}
\label{ss:relobjdist}

We start our analysis by looking at how factuality \& objectivity are distributed. Knowing how their scores are spread out is important for using them in IR, e.g. to avoid bias or saturation effects that may result from heavily skewed distributions \cite{PetersenSL16}. To get a clearer visualisation of trends, we use binning. We sort all documents by their probability of factuality \& objectivity, and we divide them into bins. We estimate the number of bins using Scott's formula \cite{Scott79}: 
\begin{equation}
M = \frac{R}{3.49 s} N^{1/3}
\end{equation}

\noindent where $M$ is the number of bins, $R$ is the range, $N$ is the number of data points, and $s$ is the sample variance. This results in: (a) 23 equal-sized bins of 6000 documents and 1 bin of the remaining 6144 documents for Disks4-5; (b) 40 equal-sized bins of 2000 documents and 1 bin of the remaining 3163 documents for ClueWeb09B. The probability of factuality \& objectivity in each bin is the mean of the factuality \& objectivity probabilities in that bin. The frequency of a probability in a bin is the sum of the frequencies of the probabilities in that bin. 

Figure \ref{fig:dist} shows the distribution of the probabilities of factuality \& objectivity in our data. We see a heavy-tailed trend for factuality: a very large number of documents have a very low (but nonzero, namely 0.04) probability of factuality, and only very few have high factuality. This trend is the same in both datasets, but the peak of the distribution is at a much higher value in ClueWeb09B ($\sim$80K) than in Disks4-5 ($\sim$ 45K). This means that the number of low factuality documents is much higher in ClueWeb09B ($\sim$80K/83K $\approx$ 94\% of all documents) than in Disks4-5 ($\sim$45K/144K $\approx$ 32\% of all documents). This may be due to the curated vs. uncurated domain difference between the collections: official or journalistic documents are overall more likely to be reliable than documents crawled from the web. 

The probability of objectivity (Figure \ref{fig:dist}) is distributed almost inversely to factuality: there seem to be more documents with higher than lower objectivity, implying that most documents are not highly biased. Note again the difference in scale of the peak of the distribution for the two datasets: the number of highly objective documents is $\sim$11K/114K $\approx$ 8\% of Disks4-5, but $\sim$80K/83K $\approx$ 94\% of ClueWeb09B. For Disks4-5 it is not surprising that only 8\% of all official/journalistic documentation is very objective, while the remaining documents have varying degrees of objectivity, without a heavy tail. Official state documentation tends to be void of personal opinions and fairly impartial, but journalistic publications, which make up most of Disks4-5, often include subjectivity \cite{Khalid12}; assuming that $p(objectivity)<0.5$ implies subjectivity, this corresponds to 0.0-0.4 on the x-axis. For ClueWeb09B however, it is surprising to see such a high rate (94\%) of very objective documents. 
This does not seem very intuitive and could be due to misestimations of the objectivity detection. To understand the causes of this, we look at the presence of spam in the ClueWeb09B subset. 

\begin{figure}
\centering
\pgfplotsset{every-axis label/.append style={font=\LARGE}}
\tikzset{every mark/.append style={font=\huge}}
\scalebox{0.9}{
\begin{tabular}{cc}
\begin{tikzpicture}[baseline,scale=0.5]
			\begin{axis}[
			font=\LARGE,
			align =center, 				
			title= {ClueWeb09B\\ Spearman's $\rho$: 0.62 ($P$-value: 0)},		
			ylabel=Spam rank \cite{Cormack11},
			xlabel=p(factuality)
			]
			\pgfplotstableread{cweb.RLB.SPM.plot}\table 
			\addplot[only marks, mark=+] table[x index=0,y index=1] from \table;
			\addplot+[smooth] coordinates{(0.03,70) (0.99,70)};
\end{axis}
\end{tikzpicture}&
\begin{tikzpicture}[baseline,scale=0.5]
			\begin{axis}[
			font=\LARGE,
			align =center, 			
			title= {ClueWeb09B\\ Spearman's $\rho$: -0.80 ($P$-value: 0)},	
			ylabel=Spam rank \cite{Cormack11},
			xlabel=p(objectivity)
			]
			\pgfplotstableread{cweb.OBJ.SPM.plot}\table 
			\addplot[only marks, mark=+] table[x index=0,y index=1] from \table;
			\addplot+[smooth] coordinates{(0.73,70) (1,70)};
\end{axis}
\end{tikzpicture}
\end{tabular}%
}
\caption{Probability of factuality \& objectivity (x-axis) vs. spam rank \cite{Cormack11} (y-axis), (binned). 
Points below the horizontal line are considered spam.}
\label{fig:spam}
\end{figure}
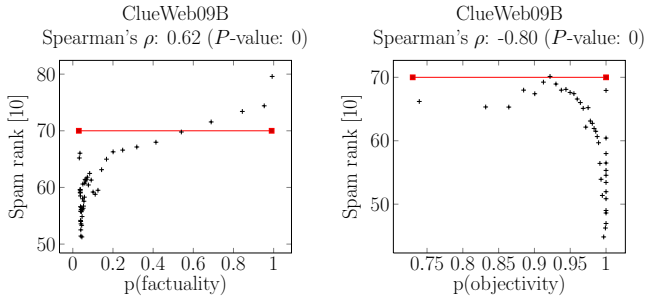

\paragraph{The effect of spam in ClueWeb09B}
We use the spam scores of \cite{Cormack11} for ClueWeb09B, where scores $>$70 are interpreted as spam. Under this interpretation, spam makes up $\sim$70\% of the whole ClueWeb09 and $\sim$49\% of our TREC-assessed subset of ClueWeb09B. Figure \ref{fig:spam} plots the probability of factuality \& objectivity in the ClueWeb09B subset vs. the spam score per bin. The same binning described earlier is used. The spam score of each bin is the mean of all spam scores in that bin. In Figure \ref{fig:spam}, left plot, we see an almost linear relation between factuality and spam, where higher factuality corresponds to lower spam, and vice versa. This makes sense intuitively; spam is notoriously unreliable. Even though the correlation between factuality and spam is not extremely strong (Spearman's $\rho$: 0.6), it is statistically significant, and the trend is visually fairly clear. The dense concentration of spam ($\sim$50-60 on the y-axis) corresponds to a low probability of factuality ($<$0.2). This indicates that most of the documents are spam and of low factuality, which agrees with our earlier finding that $\sim$49\% of this data is spam.  

For objectivity (Figure \ref{fig:spam}, right plot) we see a different trend: there is a strong negative correlation between objectivity and spam (Spearman's $\rho$: --0.8), and the relation is not diagonally linear but rather mostly `vertical', i.e. documents of high objectivity have a wide range of spam scores. Furthermore, there are very few non-spam documents: only one bin is just above the spam threshold, i.e. 2000 documents $\approx$ 2\% of the whole subset. This 2\% of non-spam tends to have high objectivity ($\sim$0.9), which seems intuitively reasonable. The remaining 98\% of the documents are spam and have various degrees of objectivity (all however are fairly highly objective, namely $>0.7$ on the x-axis, as pointed out above). Therefore most of the $\sim$80K highly objective documents in ClueWeb09B shown in Figure \ref{fig:dist} must be spam, because only $\sim$2000K of them can be non-spam. 

The strong presence of spam in the TREC-assessed subset of ClueWeb09B may impact objectivity detection more than factuality detection. The cause could be that spam contains \textit{word stuffing}: gratuitous keywords inserted to improve the retrieved rank of the document \cite{Cormack11}. 
Whereas our detection of factuality relies mainly on relations denoted by verbs (relational tuples), our detection of objectivity, while also using verb patterns, emphasises primarily noun or noun phrase lists associated with objectivity \& subjectivity (see Section \ref{ss:relsub}). \textit{Word stuffing} is typically manifested by the insertion of nouns and adjectives rather than verbs. 
Hence, it is likely that word stuffing may muddle the nominal processing of objectivity detection in ClueWeb09B, and result in artificially inflated objectivity scores. Manual inspection of the objective nouns and lexicon used for our objectivity detection confirms that this is the case.

\subsection{Factuality, Objectivity and Relevance}
\label{ss:relobjrlv}
We compare the above probabilities of factuality \& objectivity of our data to their probability of relevance: given the document bins of factuality \& objectivity described in Section \ref{ss:relobjdist}, we estimate the probability that a randomly selected relevant document belongs to a bin as: 
\begin{equation}
p(d \in b_i|relevance) = \frac{|d_{RLV}\;\in\;b_i|}{|d_{RLV}|}
\end{equation}

\noindent where $d_{RLV}$ is a relevant document according to the TREC relevance assessments, $b_i$ is the $i^{th}$ bin, and $|\cdot|$ denotes cardinality. 
We refer to $p(d \in b_i|relevance)$ as the probability of relevance in that bin.

Figure~\ref{fig:prelAnal} plots the probability of relevance (x-axis) vs. the probability of factuality or objectivity per bin (y-axis). 
We see that $p(relevance)$ varies non-randomly across bins: factuality is positively, strongly and almost linearly correlated to relevance (Spearman's $\rho$: 0.83 for Disks4-5; Spearman's $\rho$: 0.92 for ClueWeb09B). This means that boosting the ranking of higher factuality documents may boost retrieval performance, especially for ClueWeb09B where the correlation is stronger. 
The range of $p(factuality)$ is smaller for Disks4-5 (up to $\sim$0.6) than for ClueWeb09B (up to $\sim$1.0). 
On first thought this seems to indicate that there are documents of higher factuality in ClueWeb09B than in Disks4-5. Closer inspection however reveals that this is due to the different size of bins in Disks4-5 (6000 documents) and ClueWeb09B (2000 documents). 
The bigger the size of the bin, the harder to get a mean close to the maximum value of 1\footnote{Recall that $p(factuality)$ in a bin is the mean of the probabilities of factuality of all documents in that bin.}; it is more likely that the mean is `diluted' by values away from the maximum, hence it drops. 

Objectivity (Figure~\ref{fig:prelAnal}) is not correlated to relevance in Disks4-5 (Spearman's $\rho$: 0.14), and only negatively and non-linearly correlated to relevance in ClueWeb09B (Spearman's $\rho$: -0.68). 
The non-linear shape means that the lower probability of relevance corresponds to documents of both minimum and maximum probability of objectivity. Hence, using these objectivity scores for retrieval may produce mixed results. We also note that there seem to be more high objectivity documents in ClueWeb09B (most points sit above 0.9 on the y-axis) than in Disks4-5. This was already indicated in Figure \ref{fig:dist}, where  we discussed that 98\% of these documents are considered spam, meaning that measuring their objectivity is likely biased by spam effects such as \textit{word stuffing}. Hence, using our ClueWeb09B objectivity scores for retrieval is likely to produce inconclusive results.

Overall, the findings of this section are: (I) There is a strong positive correlation between relevance and factuality, especially in ClueWeb09B. It is likely that factuality may be useful for retrieval, and especially for web documents. (II) Objectivity gives a mixed picture: for ClueWeb09B, 98\% of documents, when binned by their probability of objectivity, appear to be spam, which renders their processing and subsequent analysis inconclusive; for Disks4-5, no correlation is found at all. Hence, it is likely that objectivity may not be useful for retrieval, at least  for uncurated data. (III) 
There is noise in the above estimations, particularly when data (i) includes spam, and (ii) becomes sparse (relevant documents $<<$ factual or objective documents). This affects the accuracy of the above estimations, which we should treat as indications only. 
It remains to be seen experimentally how indicative these correlations are. We do this next.


\section{Experimental Evaluation of Factuality \& Objectivity for IR}
\label{s:eval}
We experimentally study the potential usefulness of factuality \& objectivity to retrieval by treating the probability of factuality \& objectivity in a document as a type of query independent feature, such as PageRank, that we combine with a query dependent baseline. The main idea is: (i) attach a static weight to each document based on its factuality or objectivity; and (ii) combine this weight with the query dependent baseline score, to give a new score and ranking. 
There are three main approaches to combining such features and a query dependent baseline ranking \cite{Cras05}: (i) rank fusion, (ii) language modelling priors, and (iii) relevance score adjustment.
In rank fusion, the baseline and the feature scores are turned into two rankings, which are then fused, e.g. using traditional CombMNZ, voting algorithms, Bayesian inference \cite{beitzel2003}, or more recent rank interleaving methods from online learning \cite{BrostCSL16,multi}. 
In a language modelling framework, prior probabilities of the feature are calculated and combined with the language modelling probability. The third approach, and the one we use, is to rerank based on a combination of baseline and feature scores. This can be done linearly using raw scores, 
or even better with a nonlinear transformation of the input feature \cite{Cras05}. We choose to use two types of linear combination because, even though they may not give optimal performance, they make it intuitively easy to interpret the impact of the feature score on the final ranking.

\subsection{Baselines and our Methods}
Our baseline ranking model is a unigram, query likelihood, Dirichlet-smoothed, language model (DIR). We rerank the top 
1000 retrieved documents by the baseline according to their factuality or objectivity scores. Limiting the reranking to the top 1000 is more efficient than reranking all documents with a nonzero baseline score, without making a large difference to system effectiveness \cite{Cras05}. Let $S$ be the baseline ranking score of a document, which in this case is a probability but practically computed as a log for efficiency. Let $p(RLB)$ and $p(OBJ)$ be the probability of factuality \& objectivity of a document computed as per Section \ref{ss:relsub}. 
We use two reranking approaches, (i) simple \textit{linear} combination (Equation \ref{eq:linear}), and (ii) \textit{satu} \cite{Cras05} (Equation \ref{eq:satu}), to compute the reranked $S$ of each document, denoted $\widehat{S^F}$ and $\widehat{S^O}$ for factuality- and objectivity-based reranking:
\begin{equation}
\label{eq:linear}
\widehat{S^{F}_{linear}} = \log{S} \times \alpha + \log{p(FCT)} \times (1-\alpha)
\end{equation}
\noindent where $0 \le \alpha \le 1$ is a smoothing parameter controlling the effect of $S$ over $p(FCT)$. 

\begin{equation}
\label{eq:satu}
\widehat{S^{F}_{satu}} = \log{S} + w \times \frac{\log{p(FCT)}}{k+\log{p(FCT)}}     
\end{equation}
\noindent where $w,k$ are parameters: $w$ is the maximum, approached as $\log{p(FCT)}$ increases; $k$ is the value of $\log{p(FCT)}$ where \textit{satu} is $w/2$ \cite{Cras05}. 

For objectivity-based reranking ($\widehat{S^O}$), we replace $p(FCT)$ by $p(OBJ)$ in Equation \ref{eq:linear}-\ref{eq:satu}.


\subsection{Experimental Setup}
We use Indri 5.8 for indexing and retrieval without stemming or stop word removal. We use the two TREC test collections analysed in Section \ref{s:relobj} and shown in Table \ref{tab:datasets}: Disks4-5 with queries 301-450 (title only) from the AdHoc track of TREC6-8 (minus the Congressional Record for TREC7-8), and ClueWeb09B with queries 1-200 from the Web AdHoc track of TREC 2009-2012. 
We evaluate effectiveness with standard early and deep precision measures: Mean Average Precision (MAP); Normalised Discounted Cumulative Gain (NDCG); Binary Preference (BPREF), which, unlike MAP, does not treat non-assessed documents as non-relevant; Precision in top 10 (P@10); and Mean Reciprocal Rank (MRR) of the first relevant result. We test for statistical significance using the student t-test at 95\% confidence levels.

The baselines and our reranking methods include parameters that we tune using 5-fold cross-validation. We report the average of the five test folds. 
We vary DIR's $\mu \in$ \{100,500,800,1000,2000,3000,4000,5000,8000,10000\} 
and the reranking parameters $\alpha \in \{0.5..1\}$ in steps of 0.05, and $w \in \{0.5..3.5\}$ in steps of 0.5. We set $k=1$.

\subsection {Results}
\label{ss:results}

\begin{table*}
\centering
\caption{\label{tab:res.rlb}Retrieval effectiveness of the baseline (DIR), shown in grey, vs. two factuality (FCT) and objectivity (OBJ) based reranking approaches (\textit{linear} marked $\oplus$, and \textit{satu} marked $\otimes$). $\pm\%$ is the difference from the baseline. Bold means $\ge$ baseline. $\ddagger$ marks stat. significance; * marks the best score per track \& measure. 
}
\scalebox{0.75}{
\begin{tabular}{|l|lr:lr:lr|lr:lr:lr:lr|}
\hline
\multicolumn{15}{|c|}{\textbf{\large{FACTUALITY-BASED RERANKING}}}\\
\hline
\multirow{3}{*}{\bf Method}
&\multicolumn{6}{c|}{\bf CURATED DATA (Disks4-5)}&\multicolumn{8}{c|}{\bf UNCURATED DATA (ClueWeb09B)}\\
&\multicolumn{2}{c}{\bf TREC-6}	&\multicolumn{2}{c}{\bf TREC-7}	&\multicolumn{2}{c|}{\bf TREC-8	}&\multicolumn{2}{c}{\bf TREC'09}	&\multicolumn{2}{c}{\bf TREC'10}	&\multicolumn{2}{c}{\bf TREC'11}	&\multicolumn{2}{c|}{\bf TREC'12}\\
&\multicolumn{2}{c}{\bf AdHoc}	&\multicolumn{2}{c}{\bf AdHoc}	&\multicolumn{2}{c|}{\bf AdHoc}	&\multicolumn{2}{c}{\bf Web AdHoc}		&\multicolumn{2}{c}{\bf Web AdHoc}		&\multicolumn{2}{c}{\bf Web AdHoc}		&\multicolumn{2}{c|}{\bf Web AdHoc}\\
\hline
\multicolumn{15}{|c|}{\bf MAP}\\
\hline
\rowcolor{light-gray}
\bf DIR			&.1721&			&.1981*&			&.2053*&			&.1433&			&.1063&			&.0977&			&.1007&\\
\bf DIR$\oplus$FCT	&\bf.1728&+0.4\%	&.1894&--4.6\%	&.2040&--0.6\%	&\bf.1466$\ddagger$&+2.3\%	&\bf.1214$\ddagger$&+14.2\%	&\bf.1056$\ddagger$*&+8.1\%	&\bf.1034$\ddagger$& +2.7\%\\
\bf DIR$\otimes$FCT&\bf.1740*&+1.1\%	&.1863&--6.3\%	&.2050&--0.1\%	&\bf.1476$\ddagger$*&+3.0\%	&\bf.1221$\ddagger$*&+14.9\%	&\bf.1049$\ddagger$&+7.4\%	&\bf.1175$\ddagger$*&+16.7\%\\
\hline
\multicolumn{15}{|c|}{\bf NDCG}\\
\hline
\rowcolor{light-gray}
\bf DIR			&.4009&			&.4426*&			&.4454&			&.3279&			&.3020&			&.2560&			&.2795&\\
\bf DIR$\oplus$FCT	&\bf.4010&+0.0\%	&.4349&--1.8\%	&.4442&--0.3\%	&\bf.3299$\ddagger$&+0.6\%	&\bf.3175$\ddagger$&+5.1\%	&\bf.2571$\ddagger$&+0.4\%	&\bf.2932$\ddagger$&+4.9\%\\
\bf DIR$\otimes$FCT&\bf.4018*&+0.2\%	&.4318&--2.5\%	&\bf.4463*&+0.2\%	&\bf.3300$\ddagger$*&+0.6\%	&\bf.3260$\ddagger$*&+7.9\%	&\bf.2603$\ddagger$*&+1.7\%	&\bf.2982$\ddagger$*&+6.7\%\\
\hline
\multicolumn{15}{|c|}{\bf BPREF}\\
\hline
\rowcolor{light-gray}
\bf DIR			&.1966&			&.2163*&			&.2191&			&.2270&			&.2322&			&.1642&			&.2335&\\
\bf DIR$\oplus$FCT	&.1965&+0.0\%		&.2086&--3.7\%	&\bf.2197&+0.3\%	&\bf.2271$\ddagger$*&+0.0\%	&\bf.2542$\ddagger$*&+9.5\%	&\bf.1825$\ddagger$&+11.1\%	&\bf.2886$\ddagger$*&+23.6\%\\
\bf DIR$\otimes$FCT&\bf.2011*&+2.3\%	&.2105&--2.8\%	&\bf.2207*&+0.7\%	&\bf.2270$\ddagger$&+0.0\%	&\bf.2538$\ddagger$&+9.3\%	&\bf.1861$\ddagger$*&+13.3\%	&\bf.2834$\ddagger$&+21.4\%\\
\hline
\multicolumn{15}{|c|}{\bf P@10}\\
\hline
\rowcolor{light-gray}
\bf DIR			&.3180&			&.4060&			&.4000&			&.9405&			&.2585&			&.2400&			&.1980&\\
\bf DIR$\oplus$FCT	&\bf.3260*&+2.5\%	&\bf.4280*&+5.4\%	&\bf.4160*&+4.0\%	&\bf.9682$\ddagger$&+2.9\%	&\bf.3445$\ddagger$&+33.3\%	&\bf.2460$\ddagger$*&+2.5\%	&\bf.2400$\ddagger$*&+21.2\%\\
\bf DIR$\otimes$FCT&.3140&--1.3\%	&\bf.4280*&+5.4\%	&\bf.4120&+3.0\%	&\bf.9723$\ddagger$*&+3.4\%	&\bf.3805$\ddagger$*&+47.2\%	&\bf.2640$\ddagger$*&+10.0\%	&\bf.2280$\ddagger$&+15.2\%\\
\hline
\multicolumn{15}{|c|}{\bf MRR}\\
\hline
\rowcolor{light-gray}
\bf DIR			&.5800&			&.7149*&			&.6932&			&1.000*&			&.3589&			&.4096&			&.2057&\\
\bf DIR$\oplus$FCT	&\bf.6176&+6.5\%	&.7021&--1.8\%	&\bf.6948*&+0.2\%	&.9817$\ddagger$&--1.9\%	&\bf.4588$\ddagger$&+27.8\%	&\bf.4939$\ddagger$*&+20.6\%	&\bf.4218$\ddagger$*&+105.1\%\\
\bf DIR$\otimes$FCT&\bf.6400*&+10.3\%	&.6974&--2.5\%	&\bf.6941&+01\%	&\bf1.000$\ddagger$*&+0.0\%	&\bf.5048$\ddagger$*&+40.7\%	&\bf.4841$\ddagger$&+18.2\%	&\bf.4104$\ddagger$&+99.5\%\\
\hline
\hline
\multicolumn{15}{|c|}{\textbf{\large{OBJECTIVITY-BASED RERANKING}}}\\
\hline
\multirow{3}{*}{\bf Method}
&\multicolumn{6}{c|}{\bf CURATED DATA (Disks4-5)}&\multicolumn{8}{c|}{\bf UNCURATED DATA (ClueWeb09B)}\\
&\multicolumn{2}{c}{\bf TREC-6}	&\multicolumn{2}{c}{\bf TREC-7}	&\multicolumn{2}{c|}{\bf TREC-8	}&\multicolumn{2}{c}{\bf TREC'09}	&\multicolumn{2}{c}{\bf TREC'10}	&\multicolumn{2}{c}{\bf TREC'11}	&\multicolumn{2}{c|}{\bf TREC'12}\\
&\multicolumn{2}{c}{\bf AdHoc}	&\multicolumn{2}{c}{\bf AdHoc}	&\multicolumn{2}{c|}{\bf AdHoc}	&\multicolumn{2}{c}{\bf Web AdHoc}		&\multicolumn{2}{c}{\bf Web AdHoc}		&\multicolumn{2}{c}{\bf Web AdHoc}		&\multicolumn{2}{c|}{\bf Web AdHoc}\\
\hline
\multicolumn{15}{|c|}{\bf MAP}\\
\hline
\rowcolor{light-gray}
\bf DIR			&.1721&			&.1981*&			&.2053*&			&.1433&			&.1063*&			&.0977&			&.1007&\\
\bf DIR$\oplus$OBJ	&.1720&--0.1\%	&.1856&--6.7\%	&.1970&--4.2\%	&\bf.1438$\ddagger$*&+0.3\%	&\bf.1063$\ddagger$*&+0.0\%	&\bf.0980$\ddagger$*&+0.3\%	&\bf.1014$\ddagger$&+0.7\%\\
\bf DIR$\otimes$OBJ&\bf.1732*&+0.6\%	&.1864&--6.3\%	&.2039&--0.7\%	&\bf.1437$\ddagger$&+0.3\%	&.1059$\ddagger$&--0.4\%	&\bf.0977$\ddagger$&+0.0\%	&\bf.1023$\ddagger$*&+1.6\%\\
\hline
\multicolumn{15}{|c|}{\bf NDCG}\\
\hline
\rowcolor{light-gray}
\bf DIR			&.4009*&			&.4426*&			&.4454*&			&.3279&			&.3020*&			&.2560&			&.2795&\\
\bf DIR$\oplus$OBJ	&.4001&--0.2\%	&.4331&--2.2\%	&.4185&--6.4\%	&\bf.3283*&+0.1\%	&\bf.3020*&+0.0\%	&\bf.2560&+0.0\%	&.2794&+0.0\%\\
\bf DIR$\otimes$OBJ&.4008&+0.0\%	&.4313&--2.6\%	&.4447&--0.2\%	&\bf.3282&+0.1\%	&.3017&--0.1\%	&\bf.2561*&+0.0\%	&\bf.2813*&+0.6\%\\
\hline
\multicolumn{15}{|c|}{\bf BPREF}\\
\hline
\rowcolor{light-gray}
\bf DIR			&.1966&			&.2163*&			&.2191*&			&.2270&			&.2322&			&.1642*&			&.2335*&\\
\bf DIR$\oplus$OBJ	&.1963&--0.2\%		&.2087&--3.6\%	&.2132&--2.8\%	&\bf.2273$\ddagger$*&+0.1\%	&\bf.2323$\ddagger$*&+0.0\%	&\bf.1642$\ddagger$*&+0.0\%	&.2331$\ddagger$&--0.2\%\\
\bf DIR$\otimes$OBJ&\bf.2014*&+2.4\%	&.2114&--2.3\%		&.2172&--0.9\%	&\bf.2273$\ddagger$*&+0.1\%	&.2319$\ddagger$&--0.1\%	&.1638$\ddagger$&--0.2\%		&.2329$\ddagger$&--0.3\%\\
\hline
\multicolumn{15}{|c|}{\bf P@10}\\
\hline
\rowcolor{light-gray}
\bf DIR			&.3180&			&.4060&			&.4000&			&.9405*&			&.2585&			&.2400&			&.1980*&\\
\bf DIR$\oplus$OBJ	&\bf.3200*&+0.6\%	&\bf.4080&+0.5\%	&\bf.4120*&+3.0\%	&.9365$\ddagger$&--0.4\%	&\bf.2675$\ddagger$*&+3.5\%	&\bf.2440$\ddagger$*&+1.7\%	&.1940$\ddagger$&--2.1\%\\
\bf DIR$\otimes$OBJ&.3140&--1.3\%	&\bf.4240*&+4.4\%	&\bf.4100&+2.5\%	&.9365$\ddagger$&--0.4\%	&\bf.2655$\ddagger$&+2.7\%	&.2380$\ddagger$&--6.7\%	&.1940$\ddagger$&--2.1\%\\
\hline
\multicolumn{15}{|c|}{\bf MRR}\\
\hline
\rowcolor{light-gray}
\bf DIR			&.5800&			&.7149*&			&.6932*&			&1.000*&			&.3589*&			&.4096*&			&.2057&\\
\bf DIR$\oplus$OBJ	&.5681&-2.1\%		&.6921&--3.3\%	&.6913&--0.3\%	&\bf1.000$\ddagger$*&+0.0\%	&.3498$\ddagger$&--2.6\%	&.4095$\ddagger$&+0.0\%		&\bf.2185$\ddagger$&+6.2\%\\
\bf DIR$\otimes$OBJ&\bf.5863*&+0.1\%	&.6807&--5.0\%	&.6909&--0.3\%	&\bf1.000$\ddagger$*&+0.0\%	&.3492$\ddagger$&--2.8\%	&.4019$\ddagger$&--1.9\%	&\bf.2474$\ddagger$*&+20.3\%\\
\hline
\end{tabular}
}
\end{table*}
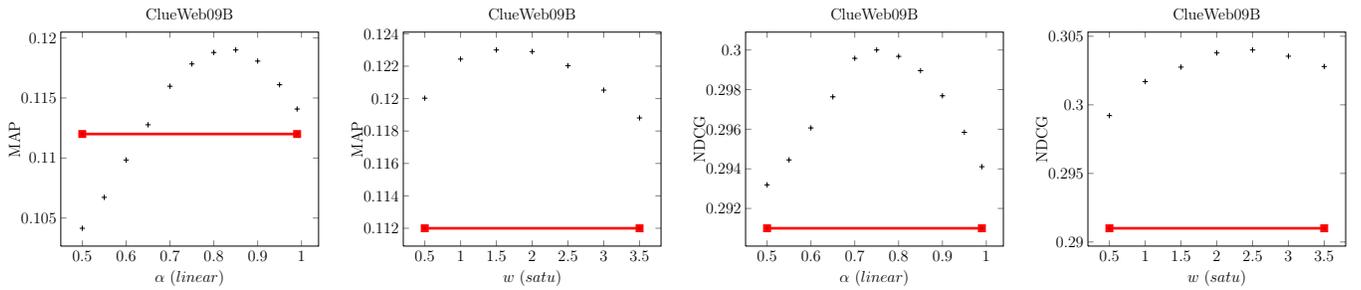
\begin{figure*}
\centering
\pgfplotsset{every-axis label/.append style={font=\large}}
\tikzset{every mark/.append style={font=\large}}
\begin{tabular}{cccc}
\begin{tikzpicture}[baseline,scale=0.5]
			\begin{axis}[
			font=\large,			
			title= ClueWeb09B,
			scaled y ticks=manual,			
			ylabel=MAP,
			xlabel=$\alpha$ ($linear$),
			y tick label style={/pgf/number format/precision=4},
			]
			\pgfplotstableread{cweb-reliab-arith-map2}\table 
			\addplot[only marks, mark=+] table[x index=0,y index=1] from \table;
			\addplot+[smooth, line width=2pt] coordinates{(0.5,0.112) (0.99,0.112)};
\end{axis}
\end{tikzpicture}&
\begin{tikzpicture}[baseline,scale=0.5]
			\begin{axis}[
			font=\large,			
			title= ClueWeb09B,
			scaled y ticks=manual,			
			ylabel=MAP,
			xlabel=$w$ ($satu$),
			y tick label style={/pgf/number format/precision=4},
			]
			\pgfplotstableread{cweb-reliab-satu-map2}\table 
			\addplot[only marks, mark=+] table[x index=0,y index=1] from \table;
			\addplot+[smooth, line width=2pt] coordinates{(0.5,0.112) (3.5,0.112)};
\end{axis}
\end{tikzpicture}&
\begin{tikzpicture}[baseline,scale=0.5]
			\begin{axis}[
			font=\large,			
			title= ClueWeb09B,
			ylabel=NDCG,
			xlabel=$\alpha$ ($linear$),
			y tick label style={/pgf/number format/precision=4},
			]
			\pgfplotstableread{cweb-reliab-arith-ndcg2}\table 
			\addplot[only marks, mark=+] table[x index=0,y index=1] from \table;
			\addplot+[smooth, line width=2pt] coordinates{(0.5,0.291) (0.99,0.291)};

\end{axis}
\end{tikzpicture}&
\begin{tikzpicture}[baseline,scale=0.5]
			\begin{axis}[
			font=\large,			
			title= ClueWeb09B,
			scaled y ticks=manual,			
			ylabel=NDCG,
			xlabel=$w$ ($satu$),
			y tick label style={/pgf/number format/precision=4},
			]
			\pgfplotstableread{cweb-reliab-satu-ndcg2}\table 
			\addplot[only marks, mark=+] table[x index=0,y index=1] from \table;
			\addplot+[smooth, line width=2pt] coordinates{(0.5,0.291) (3.5,0.291)};
\end{axis}
\end{tikzpicture}
\end{tabular}%
\caption{Factuality reranking parameter values for \textit{linear} and \textit{satu} (x-axis) vs. MAP/NDCG (y-axis) in ClueWeb09B. The horizontal line marks baseline performance.}
\label{fig:sensitivity}
\end{figure*}

For \textbf{factuality reranking} (upper part of Table \ref{tab:res.rlb}), baseline performance is generally higher for MAP and NDCG in the curated than uncurated datasets (more mixed picture for BPREF, P@10, MRR). This trend agrees with the performances recorded in past TREC proceedings and is likely due to the different characteristics and difficulties of the respective test collections: web queries (for the uncurated data) tend to be shorter and to have fewer relevance assessments in relation to the collection size, hence they are somewhat harder to satisfy than earlier Disks4-5 queries. 

For \textbf{uncurated data}, factuality reranking always outperforms the baseline, or equals it when the maximum has been reached by the baseline (e.g. for MRR), with the exception of the linear combination for the Web track 2009 where MRR slightly drops from 1.0 to 0.98. All these improvements are statistically significant and agree with the strong correlation between factuality and relevance found in Section \ref{ss:relobjrlv}. 

%

For factuality reranking in \textbf{curated data}, we see both gains, which are modest, as well as drops in performance. This could be because curated data is of higher general quality, so the distinction between low and high factuality documents is not as sharp as in uncurated data (hence there is less margin for improvement using factuality). This agrees with the weaker correlation between factuality and relevance found in Section \ref{ss:relobjrlv} for the curated dataset. 

Overall, \textit{satu} performs better than \textit{linear} most of the times, which implies that even more refined transformations of the factuality probability can potentially lead to higher performance gains.

%

For \textbf{objectivity reranking} (lower part of Table \ref{tab:res.rlb}), we see very little gain and occasional performance drops in uncurated data. Given the high amount of spam in incurated data (discussed in Section \ref{ss:relobjdist}), we consider these findings inconclusive, i.e. we refrain from drawing conclusions about the potential usefulness or not of objectivity in uncurated data retrieval. 
For curated data, objectivity gives hardly any gains at all, which agrees with the lack of correlation found between objectivity and relevance in Section \ref{ss:relobjrlv}. We conclude that objectivity reranking, as implemented here, does not seem to help curated data retrieval.  

We focus the remaining discussion on factuality and discuss some pertinent aspects of our findings.

\subsubsection{Early vs. deep precision}
Factuality improvements are overall much higher for early precision (up to +47.2\% for P@10 and  +105.1\% for MRR) than for deep precision (up to +5\% for NDCG, +16.7\% for MAP, and 23.6\% for BPREF). BPREF brings higher gains than MAP, which means that many among the retrieved documents are non-assessed and are treated blindly as non-relevant by MAP (whereas BPREF ignores them). NDCG has the lowest overall gains, probably because it punishes relevant documents at lower ranks more harshly than MAP or BPREF. This could mean that, even after factuality reranking, some relevant documents are placed lower than the top ranks. For early precision, factuality reranking seems very effective, meaning that it pushes relevant documents even higher up in the ranks 1-10. This is important because, for the Web tracks, early precision measures are possibly more important than deep precision measures \cite{Kelly10}. In addition, MAP is severely limited for the Web tracks \cite{Cormack11} because it uses in its calculation 
a binary feature $rel(k)$ of value 1 if the $k^{th}$ ranked document is relevant, otherwise 0. However, $rel(k)$ is unknown for most $k>12$ in ClueWeb09B, and methods to estimate MAP with incomplete knowledge of $rel(k)$ have proven to be unreliable \cite{Cormack11}. 

The overall gains in precision, both averaged over the top 1000 documents, and in the single first relevant and top 10 retrieved documents, mean that factuality benefits ranking across the range of relevant documents in web search (those retrieved in the top ranks and those retrieved further down). 


\begin{table}
\centering
\caption{\label{tab:perquery}Average baseline performance for queries grouped by their $\pm$\% change over the baseline when reranked by factuality ($\pm$=0\% is ignored). The number of queries is in parentheses.} 
\scalebox{0.8}{
\begin{tabular}{c| r r | r r | r r | r r } 
\rowcolor{light-gray}
$\pm$\%  &\multicolumn{4}{c|}{MAP}	&\multicolumn{4}{c}{NDCG} \\
\hline
			&\multicolumn{2}{c|}{Disks4-5}	&\multicolumn{2}{c|}{ClueWeb09B} 	 	&\multicolumn{2}{c|}{Disks4-5}	&\multicolumn{2}{c}{ClueWeb09B} 		\\
\hline
$>$10\%		&.1310 &(28)					&.1141 &(90)						&.2242 &(11)					&.3235 &(68)	\\
5.0-10\%		&.2355 &(16)					&.1626 &(20)						&.3202 &(12)					&.3084 &(11)	\\
0.1-4.9\%		&.2760 &(40)					&.1528 &(61)						&.5503 &(59)					&.3163 &(66)	\\
$<$0\%			&.2154 &(59)					&.1040 &(22)						&.4588 &(65)					&.2979 &(43)	\\
\end{tabular}
}
\end{table}

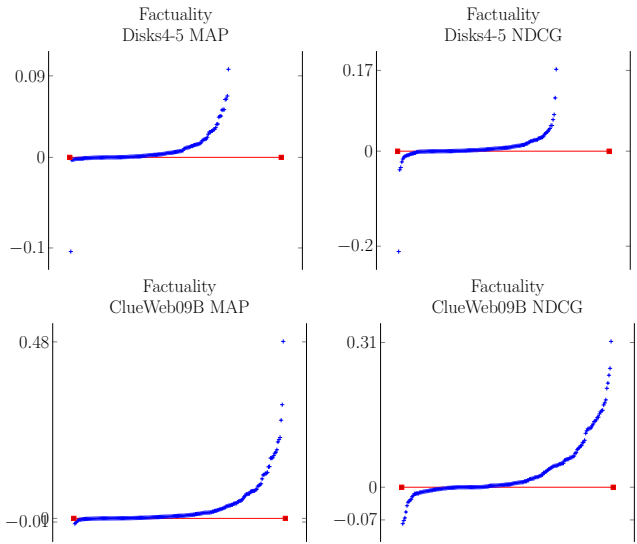
\begin{figure}
\centering
\pgfplotsset{every-axis label/.append style={font=\footnotesize}}
\tikzset{every mark/.append style={font=\scriptsize}}
\pgfplotsset{every-axis legend/.append style={
			at={(0.5,-0.0)},
			anchor=south}
			}
\scalebox{0.8}{
\begin{tabular}{cc}
\begin{tikzpicture}[baseline,scale=0.5]
\begin{axis}[
scale only axis,
ytick={-0.1,0.0,+0.09},
hide x axis=true,
font=\LARGE,
scaled y ticks = false,
y tick label style={/pgf/number format/fixed},
align =center,
title={Factuality\\ Disks4-5 MAP}
]
\pgfplotstableread{qid-mapDiff-d45}\table 
\addplot+[only marks, mark=+] table[x index=0,y index=1] from \table;
\addplot+[smooth] coordinates{(0,0) (200,0)};
\end{axis}
\end{tikzpicture} &
\begin{tikzpicture}[baseline,scale=0.5]
\begin{axis}[
scale only axis,
ytick={-0.2,0.0,+0.17},
hide x axis=true,
font=\LARGE,
scaled y ticks = false,
y tick label style={/pgf/number format/fixed},
align =center,
title={Factuality\\Disks4-5 NDCG}
]
\pgfplotstableread{qid-ndcgDiff-d45}\table 
\addplot+[only marks, mark=+] table[x index=0,y index=1] from \table;
\addplot+[smooth] coordinates{(0,0) (200,0)};
\end{axis}
\end{tikzpicture}\\
\begin{tikzpicture}[baseline,scale=0.5]
\begin{axis}[
scale only axis,
ytick={-0.01,0.0,+0.48},
hide x axis=true,
font=\LARGE,
scaled y ticks = false,
y tick label style={/pgf/number format/fixed},
align =center,
title={Factuality\\ClueWeb09B MAP}
]
\pgfplotstableread{qid-mapDiff-cweb}\table 
\addplot+[only marks, mark=+] table[x index=0,y index=1] from \table;
\addplot+[smooth] coordinates{(0,0) (200,0)};
\end{axis}
\end{tikzpicture}&
\begin{tikzpicture}[baseline,scale=0.5]
\begin{axis}[
scale only axis,
ytick={-0.07,0.0,+0.31},
hide x axis=true,
font=\LARGE,
scaled y ticks = false,
y tick label style={/pgf/number format/fixed},
align =center,
title={Factuality\\ClueWeb09B NDCG}
]
\pgfplotstableread{qid-ndcgDiff-cweb}\table 
\addplot+[only marks, mark=+] table[x index=0,y index=1] from \table;
\addplot+[smooth] coordinates{(0,0) (200,0)};
\end{axis}
\end{tikzpicture}\\
\end{tabular}
}
\caption{\label{fig:diff} Sorted per-query difference in MAP/NDCG between the baseline (DIR) and both factuality reranking methods (\textit{linear} \& \textit{satu}) (y-axis). The horizontal line marks the baseline (points above are gains). Each point is a query.}
\end{figure}

\subsubsection{Per query analysis}
Breaking down MAP and NDCG on a per query basis (Figure \ref{fig:diff}) illustrates the difference between curated and uncurated data even more clearly: for factuality reranking in curated data the gain is slightly lower than the deterioration (within -0.1 and +0.09 for MAP; within -0.2 and +0.17 for NDCG). 
For uncurated data, the gain is always higher than deterioration, the largest being within -0.01 and +0.48 for MAP. It seems that in uncurated data, there may be more potential for improvement over the baseline with factuality. Figure \ref{fig:diff} also shows that the precision gains reported in Table \ref{tab:res.rlb} are not misleadingly inflated by outliers that may affect the means of the evaluation measures, but rather they spread over most queries.


To look at which queries improve or deteriorate more closely, Table \ref{tab:perquery} groups queries by their \% difference from the baseline, and shows the average MAP and NDCG of the baseline. I.e., the row `$>$10\% .1310 (28)' means that there are 28 queries in Disks4-5 that improve by $>$10\% over the baseline, and that the average baseline MAP of these queries is .1310. We see some interesting trends: queries with low baseline MAP (i.e. harder queries) improve by $>$10\% for both curated and uncurated data; queries with higher baseline MAP (i.e. easier queries) improve by 0.1-10\% for both curated and uncurated data; the queries which deteriorate are those whose baseline is closest to the mean for curated data, and those with the lowest MAP (i.e. the hardest) for uncurated data. These trends do not hold for NDCG and curated data, where queries of higher baseline scores either gain a bit (0.1-4.9\%) or deteriorate; for uncurated data however we see the same trend as in MAP: only the hardest queries (of the lowest baseline score) deteriorate. The hardest queries are 22/150 $\approx$15\% of all Web queries. Such extremely hard queries are often very vague or ambiguous, i.e. their difficulty is due more to their formulation, than to the quality of the documents retrieved for them. One plausible explanation is thus that factuality, which is a document quality, cannot compensate for such ill-formed queries.

\subsubsection{Parameter sensitivity}
We investigate how sensitive is the effectiveness of factuality reranking with \textit{linear} and \textit{satu} to perturbations of their respective parameters $\alpha$ and $w$. If slight parameter shifts cause effectiveness to drastically change, then the respective approach is sensitive, hence more 
prone to drastic changes in effectiveness if a poor parameter setting is chosen (even though sensitive methods do not necessarily have poor generalisation properties if properly tuned \cite{Metzler06}).
Figure \ref{fig:sensitivity} shows the geometry of the metric surface of the parameters used in \textit{linear} and \textit{satu} factuality reranking for MAP and NDCG for ClueWeb09B. The metric surface of $w$ (\textit{satu}) is relatively flat across the parameter space, hence more stable than the metric surface of $\alpha$ (\textit{linear}), which has regions falling off to values of lower effectiveness. 
Similar plots are produced for factuality reranking in Disks4-5 (omitted for brevity). Hence, \textit{satu}, not only performs better, but also seems to be more stable than \textit{linear}. The stability of \textit{satu} is also reported in \cite{Cras05}.

\section{Discussion}
\label{s:disc}

\subsection{Correlation limitations}
One should be careful when interpreting the correlations presented in Section \ref{s:relobj} as a measure of association between factuality or objectivity and relevance. Correlation does not imply causation; we should not assume that a document is relevant because it is factual or objective. Nor should we assume the contrary, i.e. that there is no connection between factuality and objectivity, just because we find no statistical correlation between them (Spearman's $\rho$: -0.03, $P$-value: 0.5 for Disks4-5; Spearman's $\rho$: 0.01, $P$-value: 0.8 for ClueWeb09B). Linguistic studies find clear links between factuality and objectivity \cite{Froehlich02,OsgoodST57}.

Furthermore, some aspects of bivariate relationships are not adequately addressed by correlation, e.g. changes in scale and location, because it is a symmetrical measure \cite{Fisher58}. 
This means that effects attributable to location and differential variability, e.g. differences in importance between two variables, are not detected by correlation. 
Indeed, the scale differences between factuality/objectivity and relevance in Figure~\ref{fig:prelAnal} above are quite large: the range of $p(relevance)$ (0.01-0.05) is consistently lower than the range of $p(factuality)$ and $p(objectivity)$ (0-1). This is because the documents assessed as relevant by TREC are both scarce and sparse (known for ClueWeb09B \cite{Cormack11}), an artefact of the test collection design. In search engines where relevance assessments are simulated by features that are easier to extract in large numbers, e.g. user clicks, it is likely there may be a higher proportion of relevant documents among those assessed. Practically this may reduce the scale difference we see in Figure~\ref{fig:prelAnal}, resulting in clearer trends. 

\subsection{Efficiency}
The computational cost of detecting factuality \& objectivity for IR is not prohibitive, especially considering that it can be done off-line during indexing. For factuality detection, the most expensive step is fact extraction with ReVerb ($\sim$100ms per document on a Intel(R) Xeon(R) CPU with a 2.00GHz processor). 
 For objectivity detection, the most expensive step is parsing with the StanfordParser (20s per document on average, on an Intel(R) Xeon(R) CPU with a 2.50GHz processor). For both factuality \& objectivity, the exact computational complexity of LIBSVM is not known, but it is not linear \cite{Hsu03}. These costs can be reduced considerably by parallelising the processing or improving the efficiency of the NLP code. 

\section{Conclusion}
We presented a study of the relation between document factuality \& objectivity and document relevance for IR. We used state-of-the-art factuality \& objectivity detection from NLP on two TREC collections, one curated (official documentation/newswire) and one uncurated (web crawl). We found that (i) factuality is positively correlated to relevance, and the correlation is stronger for uncurated data; (ii) objectivity is not correlated to relevance for curated data, while, for uncurated data, the high presence of spam hinders its detection. We further experimented with factuality \& objectivity as document scores, which we combined linearly with competitive baseline scores to rerank the top 1000 search results of 350 TREC queries. Factuality was found to be highly effective, especially for uncurated data, gaining $>$10\% in average precision. Objectivity findings were mixed and inconclusive. 

To our knowledge, this is the first study of document factuality \& objectivity for back-end IR. 
Next steps include repeating the objectivity analysis on spam-free (or spam-light) uncurated data. We did not attempt this for ClueWeb09B because removing its spam would also remove ca. half of its TREC assessed documents, leaving us with very few documents to study ($\sim$42K or 0.08\% of the collection). 
Another 
interesting future research direction is to explore how best to incorporate factuality and objectivity document scores to retrieval, e.g. by replacing the linear score combinations we used here for reranking with supervised learning for computing optimal rerankings, given the original ranking and the factuality or objectivity scores.

\section{Acknowledgment}
Work partially funded by C. Lioma's \textit{FREJA research excellence} fellowship (grant no. 790095), and W. Lu's National Natural Science Foundation of China grant (no. 71173164).
%
%
%
\bibliographystyle{abbrv}
\bibliography{sig-powerproc}  
%
%

\end{document}